\batchmode
\documentstyle[12pt]{article}
\newcommand{\nc}{\newcommand}
\nc{\renc}{\renewcommand}

\nc{\half}{{\textstyle{1\over2}}}
\nc{\etal}{\mbox{\it et al. }}
\nc{\ie}{{\it i.e.}}
\nc{\eg}{{\it e.g.}}

\renc{\thefootnote}{\arabic{footnote}}
\nc{\capt}[1]{{\bf Figure.} {\small\sl #1}}

\nc{\eqs}[2]{\mbox{Eqs.~(\ref{#1},\,\ref{#2})}}
\nc{\eq}[1]{\mbox{Eq.~(\ref{#1})}}

\nc{\figs}[2]{\mbox{Figs.~(\ref{#1},\,\ref{#2})}}
\nc{\fig}[1]{\mbox{Fig~.(\ref{#1})}}

\nc{\tag}[1]{\label{#1} \marginpar{{\footnotesize #1}}}
\nc{\mtag}[1]{\label{#1} \mbox{\marginpar{{\footnotesize #1}}}}
\renc{\baselinestretch}{1.5}
\newlength{\overeqskip}
\newlength{\undereqskip}
\nc{\be}[1]{\begin{equation} \mbox{$\label{#1}$}}
\nc{\bea}[1]{\begin{eqnarray} \mbox{$\label{#1}$}}
\nc{\Section}[2]{\section{#2}\label{#1}}
\nc{\Bibitem}[1]{\bibitem{#1}}
\nc{\Label}[1]{\label{#1}}
\nc{\eea}{\vspace{\undereqskip}\end{eqnarray}}
\nc{\ee}{\vspace{\undereqskip}\end{equation}}
\nc{\bdm}{\begin{displaymath}}
\nc{\edm}{\end{displaymath}}
\nc{\dpsty}{\displaystyle}
\nc{\bc}{\begin{center}}
\nc{\ec}{\end{center}}
\nc{\ba}{\begin{array}}
\nc{\ea}{\end{array}}
\nc{\bab}{\begin{abstract}}
\nc{\eab}{\end{abstract}}
\nc{\btab}{\begin{tabular}}
\nc{\etab}{\end{tabular}}
\nc{\bit}{\begin{itemize}}
\nc{\eit}{\end{itemize}}
\nc{\ben}{\begin{enumerate}}
\nc{\een}{\end{enumerate}}
\nc{\bfig}{\begin{figure}}
\nc{\efig}{\end{figure}}
\nc{\arreq}{&\!=\!&}
\nc{\arrmi}{&\!-\!&}
\nc{\arrpl}{&\!+\!&}
\nc{\arrap}{&\!\!\!\approx\!\!\!&}
\nc{\non}{\nonumber\\*}
\nc{\align}{\!\!\!\!\!\!\!\!&&}

\def\lsim{\; \raise0.3ex\hbox{$<$\kern-0.75em
      \raise-1.1ex\hbox{$\sim$}}\; }
\def\gsim{\; \raise0.3ex\hbox{$>$\kern-0.75em
      \raise-1.1ex\hbox{$\sim$}}\; }
\nc{\DOT}{\hspace{-0.08in}{\bf .}\hspace{0.1in}}
\nc{\Laada}{\hbox {$\sqcap$ \kern -1em $\sqcup$}}
\nc\loota{{\scriptstyle\sqcap\kern-0.55em\hbox{$\scriptstyle\sqcup$}}}
\nc\Loota{{\sqcap\kern-0.65em\hbox{$\sqcup$}}}
\nc\laada{\Loota}
\nc{\qed}{\hskip 3em \hbox{\BOX} \vskip 2ex}

\nc{\real}{{\rm I \! R}}
\nc{\Z}{{\sf Z \!\!\! Z}}
\nc{\complex}{{\rm C\!\!\! {\sf I}\,\,}}
\def\bigid{\leavevmode\hbox{\small1\kern-3.8pt\normalsize1}}
\def\id{\leavevmode\hbox{\small1\kern-3.3pt\normalsize1}}
\nc{\slask}{\!\!\!/}
\nc{\bis}{{\prime\prime}}
\nc{\pa}{\partial}
\nc{\na}{\nabla}
\nc{\ra}{\rangle}
\nc{\la}{\langle}
\nc{\goto}{\rightarrow}
\nc{\swap}{\leftrightarrow}

\nc{\EE}[1]{ \mbox{$\cdot10^{#1}$} }
\nc{\abs}[1]{\left|#1\right|}
\nc{\at}[2]{\left.#1\right|_{#2}}
\nc{\norm}[1]{\|#1\|}
\nc{\abscut}[2]{\Abs{#1}_{\scriptscriptstyle#2}}
\nc{\vek}[1]{{\rm\bf #1}}
\nc{\integral}[2]{\int\limits_{#1}^{#2}}
\nc{\inv}[1]{\frac{1}{#1}}
\nc{\dd}[2]{{{\partial #1}\over{\partial #2}}}
\nc{\ddd}[2]{{{{\partial}^2 #1}\over{\partial {#2}^2}}}
\nc{\dddd}[3]{{{{\partial}^2 #1}\over
	{\partial #2 \partial #3}}}
\nc{\dder}[2]{{{d #1}\over{d #2}}}
\nc{\ddder}[2]{{{d^2 #1}\over{d {#2}^2}}}
\nc{\dddder}[3]{{d^2 #1}\over
	{d #2 d #3}}
\nc{\dx}[1]{d\,^{#1}x}
\nc{\dy}[1]{d\,^{#1}y}
\nc{\dz}[1]{d\,^{#1}z}
\nc{\dl}[1]{\frac{d\,^{#1}l}{(2\pi)^{#1}}}
\nc{\dk}[1]{\frac{d\,^{#1}k}{(2\pi)^{#1}}}
\nc{\dq}[1]{\frac{d\,^{#1}q}{(2\pi)^{#1}}}

\nc{\cc}{\mbox{$c.c.$ }}
\nc{\hc}{\mbox{$h.c.$ }}
\nc{\cf}{cf.\ }
\nc{\erfc}{{\rm erfc}}
\nc{\Tr}{{\rm Tr\,}}
\nc{\tr}{{\rm tr\,}}
\nc{\pol}{{\rm pol}}
\nc{\sign}{{\rm sign}}
\nc{\bfT}{{\bf T }}

\def\GeV{{\rm\ GeV}}
\def\MeV{{\rm\ MeV}}

\nc{\cA}{{\cal A}}
\nc{\cB}{{\cal B}}
\nc{\cD}{{\cal D}}
\nc{\cE}{{\cal E}}
\nc{\cG}{{\cal G}}
\nc{\cH}{{\cal H}}
\nc{\cL}{{\cal L}}
\nc{\cO}{{\cal O}}
\nc{\cT}{{\cal T}}
\nc{\cN}{{\cal N}}
\nc{\rvac}[1]{|{\cal O}#1\rangle}
\nc{\lvac}[1]{\langle{\cal O}#1|}
\nc{\rvacb}[1]{|{\cal O}_\beta #1\rangle}
\nc{\lvacb}[1]{\langle{\cal O}_\beta #1 |}
\nc{\bb}{\bar{\beta}}
\nc{\bt}{\tilde{\beta}}
\nc{\ctH}{\tilde{\cal H}}
\nc{\chH}{\hat{\cal H}}
\nc{\1}{\aa}
\nc{\2}{\"{a}}
\nc{\3}{\"{o}}
\nc{\4}{\AA}
\nc{\5}{\"{A}}
\nc{\6}{\"{O}}
\nc{\al}{\alpha}
\nc{\g}{\gamma}
\nc{\Del}{\Delta}
\nc{\e}{\epsilon}
\nc{\eps}{\epsilon}
\nc{\lam}{\lambda}
\nc{\om}{\omega}
\nc{\Om}{\Omega}
\nc{\ve}{\varepsilon}
\nc{\mn}{{\mu\nu}}
\nc{\k}{\kappa}
\nc{\vp}{\varphi}

\nc{\advp}[3]{{\it  Adv.\ in\ Phys.\ }{{\bf #1} {(#2)} {#3}}}
\nc{\annp}[3]{{\it  Ann.\ Phys.\ (N.Y.)\ }{{\bf #1} {(#2)} {#3}}}
\nc{\apl}[3]{{\it  Appl. Phys. Lett. }{{\bf #1} {(#2)} {#3}}}
\nc{\apj}[3]{{\it  Ap.\ J.\ }{{\bf #1} {(#2)} {#3}}}
\nc{\apjl}[3]{{\it  Ap.\ J.\ Lett.\ }{{\bf #1} {(#2)} {#3}}}
\nc{\app}[3]{{\it Astropart.\ Phys.\ }{{\bf #1} {(#2)} {#3}}}
\nc{\cmp}[3]{{\it  Comm.\ Math.\ Phys.\ }{{ \bf #1} {(#2)} {#3}}}
\nc{\cqg}[3]{{\it  Class.\ Quant.\ Grav.\ }{{\bf #1} {(#2)} {#3}}}
\nc{\epl}[3]{{\it  Europhys.\ Lett.\ }{{\bf #1} {(#2)} {#3}}}
\nc{\ijmp}[3]{{\it Int.\ J.\ Mod.\ Phys.\ }{{\bf #1} {(#2)} {#3}}}
\nc{\ijtp}[3]{{\it Int.\ J.\ Theor.\ Phys.\ }{{\bf #1} {(#2)} {#3}}}
\nc{\jmp}[3]{{\it  J.\ Math.\ Phys.\ }{{ \bf #1} {(#2)} {#3}}}
\nc{\jpa}[3]{{\it  J.\ Phys.\ A\ }{{\bf #1} {(#2)} {#3}}}
\nc{\jpc}[3]{{\it  J.\ Phys.\ C\ }{{\bf #1} {(#2)} {#3}}}
\nc{\jap}[3]{{\it J.\ Appl.\ Phys.\ }{{\bf #1} {(#2)} {#3}}}
\nc{\jpsj}[3]{{\it J.\ Phys.\ Soc.\ Japan\ }{{\bf #1} {(#2)} {#3}}}
\nc{\lmp}[3]{{\it Lett.\ Math.\ Phys.\ }{{\bf #1} {(#2)} {#3}}}
\nc{\mpl}[3]{{\it  Mod.\ Phys.\ Lett.\ }{{\bf #1} {(#2)} {#3}}}
\nc{\ncim}[3]{{\it  Nuov.\ Cim.\ }{{\bf #1} {(#2)} {#3}}}
\nc{\np}[3]{{\it  Nucl.\ Phys.\ }{{\bf #1} {(#2)} {#3}}}
\nc{\npps}[3]{{\it  Nucl.\ Phys.\ Proc.\ Suppl.\ }{{\bf #1} {(#2)} {#3}}}
\nc{\pr}[3]{{\it Phys.\ Rev.\ }{{\bf #1} {(#2)} {#3}}}
\nc{\pra}[3]{{\it  Phys.\ Rev.\ A\ }{{\bf #1} {(#2)} {#3}}}
\nc{\prb}[3]{{\it  Phys.\ Rev.\ B\ }{{{\bf #1} {(#2)} {#3}}}}
\nc{\prc}[3]{{\it  Phys.\ Rev.\ C\ }{{\bf #1} {(#2)} {#3}}}
\nc{\prd}[3]{{\it  Phys.\ Rev.\ D\ }{{\bf #1} {(#2)} {#3}}}
\nc{\prl}[3]{{\it Phys.\ Rev.\ Lett.\ }{{\bf #1} {(#2)} {#3}}}
\nc{\pl}[3]{{\it  Phys.\ Lett.\ }{{\bf #1} {(#2)} {#3}}}
\nc{\prep}[3]{{\it Phys.\ Rep.\ }{{\bf #1} {(#2)} {#3}}}
\nc{\prsl}[3]{{\it Proc.\ R.\ Soc.\ London\ }{{\bf #1} {(#2)} {#3}}}
\nc{\ptp}[3]{{\it  Prog.\ Theor.\ Phys.\ }{{\bf #1} {(#2)} {#3}}}
\nc{\ptps}[3]{{\it  Prog\ Theor.\ Phys.\ suppl.\ }{{\bf #1} {(#2)} {#3}}}
\nc{\physa}[3]{{\it  Physica\ A\ }{{\bf #1} {(#2)} {#3}}}
\nc{\physb}[3]{{\it  Physica\ B\ }{{\bf #1} {(#2)} {#3}}}
\nc{\phys}[3]{{\it Physica\ }{{\bf #1} {(#2)} {#3}}}
\nc{\rmp}[3]{{\it  Rev.\ Mod.\ Phys.\ }{{\bf #1} {(#2)} {#3}}}
\nc{\rpp}[3]{{\it Rep.\ Prog.\ Phys.\ }{{\bf #1} {(#2)} {#3}}}
\nc{\sjnp}[3]{{\it Sov.\ J.\ Nucl.\ Phys.\ }{{\bf #1} {(#2)} {#3}}}
\nc{\spjetp}[3]{{\it Sov.\ Phys.\ JETP\ }{{\bf #1} {(#2)} {#3}}}
\nc{\yf}[3]{{\it Yad.\ Fiz.\ }{{\bf #1} {(#2)} {#3}}}
\nc{\zetp}[3]{{\it Zh.\ Eksp.\ Teor.\ Fiz.\  }{{\bf #1}  {(#2)} {#3}}}
\nc{\zp}[3]{{\it Z.\ Phys.\ }{{\bf #1} {(#2)} {#3}}}
\nc{\ibid}[3]{{\sl ibid.\ }{{\bf #1} {#2} {#3}}}
\nc{\rf}[1]{(\ref{#1})}
\nc{\nn}{\nonumber \\*}
\nc{\bfB}{\bf{B}}
\nc{\bfv}{\bf{v}}
\nc{\bfx}{\bf{x}}
\nc{\bfy}{\bf{y}}
\nc{\vx}{\vec{x}}
\nc{\vy}{\vec{y}}
\nc{\oB}{\overline{B}}
\nc{\oI}{\overline{I}}
\nc{\oR}{\overline{R}}
\nc{\rar}{\rightarrow}
\nc{\ti}{\times}
\nc{\slsh}{\hskip-5pt/}
\nc{\sm}{Standard~Model~}
\nc{\MP}{M_{\rm Pl}}
\nc{\tp}{t_{\rm Pl}}
\nc{\ave}{\bar{E}}

\nc{\eff}{{\rm eff}}
\nc{\kk}{\vek{k}}
\nc{\pp}{{\rm p}}
\nc{\ga}{g_{a\gamma}}
\nc{\vv}{\\}
\nc{\eee}{{\bf E}}
\nc{\bbb}{{\bf B}}
\nc{\qcd}{T_{\rm QCD}}
\nc{\G}{\rm \ G}
\def\vec#1{{\bf #1}}
\def\lae{\;^{<}_{\sim} \;} \def\gae{\; ^{>}_{\sim} \;} 

\begin{document}
{\title{\vskip-2truecm{\hfill {{\small \\
	\hfill \\
	}}\vskip 1truecm}
{\LARGE  Q-ball Evolution, B-ball Baryogenesis and the $f_{B}$ Problem}}
%\vspace{1.2cm}
{\author{
{\sc  John McDonald$^{1}$}\\
{\sl\small c$ \backslash $o Dept. of Physics and Astronomy, 
University of Glasgow, 
Glasgow G12 8QQ, SCOTLAND}
}
\maketitle
%\vspace{1cm}
%\newpage
\begin{abstract}
\noindent

          We consider some of the issues surrounding the formation and evolution of 
Q-balls in the MSSM and its extensions. 
The ratio of the baryon number packed into Q-balls to that outside, 
$f_{B}$, plays a fundamental role in determining the relationship of dark matter to the
 baryon number in the Universe at present. The final value of $f_{B}$ will depend 
 upon the details of the formation of Q-balls from the collapse of the Affleck-Dine
 condensate and upon the subsequent evolution of the ensemble of Q-balls. We discuss the
 implications for neutralino dark matter in the gravity-mediated scenario and show that a
 light neutralino is necessary in most cases to account for the baryon to dark matter ratio, with
 an NMSSM singlino LSP of mass $m_{\chi} \lae 20 \GeV$ being a favoured candidate.

\end{abstract}
\vfil
\footnoterule
{\small $^1$mcdonald@physics.gla.ac.uk}

\thispagestyle{empty}
\newpage
\setcounter{page}{1}

%%%%%%%%%%%%%%%%%%%%%%%%%%%%%%%%%%%%%%%%%%%%
%%%%%%%%%%%%%%%%%%%%%%%%%%%%%%%%%%
%%%%%%%%%%%%%%%%%%%%%%%%%%%%%%%%%%%%%%%%%%%
%%%%%%%%%%%%%%%%%%%%%%%%%%%%%%%%%%%

\section{Introduction} 

                  The minimal supersymmetric (SUSY) standard model (MSSM) \cite{nilles}
 and its extensions offer a number
 of possibilities for the origin of 
the baryon asymmetry. In the MSSM, the most natural possibilities are electroweak 
baryogenesis \cite{ewb} and Affleck-Dine (AD) baryogenesis \cite{ad}.
 If we include the possibility of massive Majorana
neutrinos we can also have leptogenesis \cite{lg}. 

                   AD baryogenesis rather generally results in the formation of Q-balls of baryon 
number (also known as B-balls) in the early Universe [5-11].
 This occurs because the scalar 
potential of the squarks generally has attractive interaction terms \cite{asko}, making the 
homogeneous Bose condensate of squarks unstable with respect to spatial perturbations. 
As a result, the condensate fragments to eventually form Q-balls.

                In the conventional gravity-mediated MSSM, AD baryogenesis may be 
characterized by the dimension $d$ of the non-renormalizable terms responsible for lifting 
the flat directions \cite{drt}. For R-parity ($R_{p}$) 
conserving models (necessary to 
eliminate renormalizable $B$ and $L$ violating operators
 \cite{nilles}), the dimension must be even. It 
has recently been shown that the $d=4$ models
 are generally ruled out by the effect 
of thermal corrections to the AD potential \cite{allah}, which suppresses 
the baryon number.
 Thus $d=6$ models are favoured. For CP 
violating phases $\delta_{CP}$ (expected to be of the order of 1)
 these must have a reheating temperature 
$T_{R} \approx 1 \GeV /\delta_{CP}$ in order to
 account for the observed baryon asymmetry \cite{jnu}. (Such low reheating temperatures
can be a natural feature of SUSY inflation models \cite{bbd}.)
 The resulting
 Q-balls are very long lived, typically to temperatures $1 \MeV$ - 
$1 \GeV$ \cite{bbb2}. As a result, it is possible
 that the late decay of Q-balls to baryons and neutralinos 
can explain the remarkable 
similarity of the {\it number} densities of baryons and dark matter
 particles for the case of 
WIMP dark matter with mass $O(m_{W})$, for example neutralinos \cite{bbb1,bbb2}. For $\Omega = 1$ and $\Omega_{DM} \approx 0.4$, the ratio of the number density of 
baryons to the number density of dark matter particles is given by
$n_{B}/n_{DM} \approx (1.5-7.3)\;m_{DM}/m_{W}$ \cite{jrev}. (In the following we 
refer to this as the baryon to dark matter ratio.) This is a major motivation for 
the idea of late-decaying Q-balls as a simultaneous 
source of the baryon asymmetry and the dark matter density.
More generally, late-decaying Q-balls will provide a source of non-thermal
 WIMP dark matter, resulting in different predictions for the dark matter
 density as a function of the parameters of the MSSM \cite{bbb2,bbbdm}. 
This variant of AD baryogenesis, with late-decaying Q-balls carrying baryon number, 
is known as B-Ball Baryogenesis \cite{bbb2}. 

          In the case of gauge-mediated SUSY breaking [5-9], where Q-ball formation
from the AD condensate was first proposed \cite{ks0,ks1,ks}, the potential of the 
AD scalar becomes essentially completely flat for scalar expectation values 
larger than the mass of the messenger fields which transmit SUSY breaking
\cite{ks1}. As a result, the mass of the AD scalars in the Q-balls satisfies 
$m \propto Q^{-1/4}$, and for large enough Q-ball charge the Q-balls are completely
stable with respect to decay to nucleons and may account for
 dark matter \cite{ks,ls}, with interesting experimental and astrophysical
 consequences \cite{ks2}. The cosmology of AD condensate fragmentation 
in models of gauge-mediated SUSY breaking
 will be significantly different from the case of
 gravity-mediated SUSY breaking, 
but we expect aspects of the gravity-mediated
 case to be relevant to the gauge-mediated
 scenario. In this paper we will focus on the more
 conventional gravity-mediated SUSY breaking scenario. 

              In both the gravity- and gauge-mediated scenarios the final baryon to 
dark matter ratio is crucially dependent upon the ratio of baryon number
 trapped in the Q-balls to that outside, $f_{B}$.
 Both scenarios require that $f_{B} < 1$.  
Experimental limits from LEP and Tevatron require that 
the MSSM lightest SUSY partner (LSP) neutralino mass, $m_{\chi}$, satisfies $m_{\chi}
 > 46 \GeV$ \cite{oliveetc}. 
(This assumes universal A-terms and gaugino masses at the unification scale.)  As we will show, this implies that $f_{B} \lae 0.4 (\Omega_{\chi}/0.4)$ when the Q-balls
 decay. In the
 gauge-mediated scenario, $f_{B} = 1$ would result in 
all the baryon number today being in the form of stable Q-balls, an obviously unacceptable 
scenario. 

                There have been some attempts to estimate $f_{B}$ by studying the classical
 dynamics 
of the collapse of the AD condensate to non-topological solitions. Enqvist and McDonald
 \cite{dyn}
considered the collapse of a single spherically symmetric condensate lump. It was 
observed that the value of $f_{B}$ depended upon the charge density of the original AD 
condensate relative to the maximum possible charge, which we denote by 
 $Q$. As we will show, if $Q$ is less than one then not all the energy in the original AD 
condensate can be accomodated in the form of positively charged Q-balls. One possibility
 is that the condensate collapses to higher 
energy objects, Q-axitons \cite{dyn}, which are classically stable
 but which can in principle decay to eventually reach a Q-ball. For Q-axitons it was found
 that $f_{B}$ could be 
significantly less than one if $Q \ll 1$, for example $Q=0.01$ resulted in $f_{B}$ as low
 as 0.3 \cite{dyn}.  On 
the other hand, in a lattice simulation of condensate
 fragmentation, Kasuya and Kawasaki \cite{kk}
 did 
not find evidence for $f_{B}$ significantly less than one. In 
addition, they also pointed out that it was possible to form positive and negative charged 
Q-balls, which could in principle allow all the energy in a $Q < 1$ condensate to be
 entirely
 accomodated in 
Q-balls. 

               In this paper we will discuss various aspects of the $f_{B}$ problem in more
 detail. The question of whether $ \pm$Q-balls or Q-axitons form will be seen to depend
 upon the 
perturbation of the phase of the AD condensate; if the perturbation of the phase of the 
condensate is larger than the average phase when the condensate fragments, then
 $\pm$Q-balls will tend to form; if not, then Q-axitons will form. In the case where Q-axitons form 
there is the queston of how the classically quasi-stable Q-axitons evolve to lower energy
 Q-balls. We will suggest that this might occur classically via a slow emission of energy in
 scalar field waves or via annihilations of scalars within the Q-axiton. If the excess energy 
in the Q-axiton relative to the Q-ball 
is radiated before the Q-axiton decays and before thermal neutralinos
 freeze-out then it is possible that $f_{B}$ can be small enough to account for the 
baryon to dark matter ratio for a wide range of neutralino masses.

       A very different picture emerges if the $Q < 1$ AD condensate collapses to
 $\pm$Q-balls. In this case, when the Q-balls decay, assuming $f_{B} \approx 1$ for Q-balls as
 implied by
 numerical simulations, the decaying 
$\pm$Q-balls will produce oppositely charged baryon number but contribute equally
to the density of neutralinos, resulting in a large number of neutralinos relative to 
baryons for $Q \ll 1$. (A similar situation arises if Q-axitons do not lose their excess
 energy before they decay.) We will show that it is then essential that the MSSM
neutralinos annihilate with each other to give an acceptable density. As a result, 
there will be no direct relationship between the number of baryons and of dark 
matter particles. However, we will still have a non-thermal relic density of MSSM
 neutralinos from
 Q-ball decay, which is an interesting prediction in itself. (There may be some evidence for 
non-thermal dark matter from non-singular galactic halos \cite{ntdm}.) 

          If we have a maximally charged AD condensate, $Q = 1$, then there will be
direct formation of Q-balls without a Q-axiton stage. In this case 
$f_{B} \approx 1$ and MSSM neutralinos consistent
 with experimental constraints cannot account for dark matter directly 
via Q-ball decay. Annihilations must reduce the number of 
neutralinos, again losing the direct connection with the baryon number density although
 still producing non-thermal neutralino dark matter.

    Thus it will be seen that it is generally difficult for the baryon to dark matter ratio to be
 explained via Q-ball decay in the context of the MSSM, with the only possible exception
 being the limiting case where Q-axitons form and evolve to Q-balls before neutralinos
 freeze out. However, extensions of the 
MSSM with weaker lower bounds on the LSP mass, in particular the next-to-minimal
 SUSY standard model (NMSSM) \cite{nmssm} with light singlino LSPs
 \cite{sing,stephan,klap}, will be seen to be consistent with the 
baryon to dark matter ratio.

      The paper is organized as follows. In Section 2 we consider the classical value of
 $f_{B}$ from condensate fragmentation. 
We consider the initial conditions for the linear evolution of the perturbations and the 
conditions for the formation of Q-axitons versus $\pm$Q-balls. We then comment on 
the numerical simulations of condensate fragmentation and point out that a sufficiently
 long 
evolution will be necessary to reach the time when absorption/re-emission of scalar field 
waves from collapsing condensate lumps becomes negligible
 and the condensate lump can reach their 
equilibrium states, allowing $f_{B}$ to be extracted. In Section 3 we discuss the 
 possible evolution of a Q-axiton to a stable Q-ball.
 In Section 4 we consider the consequences for neutralino dark matter in the MSSM and
 the NMSSM. In Section 5 we comment on the effects of more complicated aspects of
 realistic condensate fragmentation on the value of $f_{B}$. In Section 6 we present our
 conclusions. 

\section{Classical Evolution of the AD Condensate} 

           In this section we consider some aspects of the classical evolution of perturbations. 
The formation of Q-axitons versus $\pm$Q-balls will depend upon the initial size of the
 perturbation in the phase of the AD condensate when coherent
 oscillations begin and the baryon  asymmetry is established at $H \approx m$, 
where $m$ is the mass of the AD scalars. 

\subsection{Energy in Q-balls for a non-maximally charged condensate}

            We first show that for a non-maximally charged condensate
in the gravity-mediated scenario it is not possible
 for all the energy in the condensate to be accomodated in 
positively charged Q-balls. The energy density in a coherently oscillating scalar field
 condensate with amplitude $\phi_{o}$ is $\rho = m^{2} \phi_{o}^{2}$. The maximum
 possible charge density in the condensate
 is $\rho_{Q\;max} = \rho/m \equiv m \phi_{o}^{2}$, corresponding to the
 case where all the scalars in the condensate are charged. 
The maximum energy density in Q-balls is
 $\rho_{q-balls} = m \; \rho_{Q} = m Q\; \rho_{Q\;max} = Q \rho$, since
 the mass of the scalars in the Q-balls is to a good approximation $m$ in the
 gravity-mediated scenario \cite{bbb2} and we assume that all the charge is packed into 
Q-balls. So the energy density in Q-balls to the energy density in the condensate is 
\be{e0} \frac{\rho_{q-ball}}{\rho} = Q          ~.\ee 
Thus for a non-maximally charged condensate with $Q < 1$ not
 all the energy density can be accomodated in positively charged Q-balls.
 One must either form higher energy non-topological solitons (Q-axitons)
or have an ensemble of positively and negatively charged Q-balls.

\subsection{Conditions for Q-axiton versus $\pm$Q-ball formation}

          We next review the evolution of perturbations at $H > m$ \cite{dyn}. Perturbations of the AD 
condensate may be expected to arise from quantum fluctuations of the AD scalar during 
inflation, with 
$\delta \phi \approx H/2 \pi$. The subsequent evolution of the perturbations will 
depend upon the paramaters of the scalar potential. In general, the perturbations satisfy an
equation of the form \cite{dyn}
\be{e1} \delta \ddot{\phi} + 3 H \delta \dot{\phi} - \nabla^2 \delta \phi = - k H^2 \delta 
\phi       ~,\ee
where $k$ is determined by the parameters $c$, $|a_{\lambda}|$ and $d$ of the AD 
potential \cite{dyn}, 
\be{e2} U(\Phi) \approx (m^{2} - c H^{2})\left(1 +  K \log\left( 
\frac{|\Phi|^{2}}{M^{2}}
 \right) \right) |\Phi|^{2} 
+ \frac{\lambda^{2}|\Phi|^{2(d-1)}
}{M_{*}^{2(d-3)}} + \left( \frac{A_{\lambda} 
\lambda \Phi^{d}}{d M_{*}^{d-3}} + h.c.\right)    ~.\ee
Here $K < 0$ is due to radiative corrections (with $|K| \approx 0.01 -0.1$)
 \cite{bbb1,bbb2,asko}, $d$ is the dimension of the non-renormalizable term in the
 superpotential 
which lifts the flat direction (with $d=6$ favoured), 
$c H^{2}$ gives the order $H^2$ correction to the scalar mass \cite{h2} (with $c$
 positive and
typically of the order of one for AD scalars) 
and we assume that the natural scale of the 
non-renormalizable terms is the supergravity
mass scale $M_{*} = M_{Pl}/\sqrt{8 \pi}$ \cite{nilles}.
 The A-term also receives order $H$ corrections, 
$A_{\lambda} = A_{\lambda\;o} + a_{\lambda}H$, where $A_{\lambda\;o}$ is the 
gravity-mediated soft SUSY breaking term and $a_{\lambda}$ depends on the nature of
 the inflation model; 
for F-term inflation $|a_{\lambda}|$ is typically of the order of one \cite{drt} whilst for
 minimal D-term inflation models it is zero \cite{kmr}. 

                               For the case of the AD scalar, 
$\Phi = (\phi_{1} + i \phi_{2})/\sqrt{2}$, we find different values
 for $k$, $k_{1}$ and
 $k_{2}$, corresponding to the equations of motion for
 $\delta \phi_{1}$ and $\delta \phi_{2}$ \cite{dyn}. Thus the evolution of the $\delta
 \phi_{1}
$ and $\delta \phi_{2}$ perturbations can be quite different.
 As a result, the initial $\delta \phi_{1}$
 and $\delta \phi_{2}$ at $H \approx m$ depends upon the parameters of the potential.
In particular, if $|a_{\lambda}|$ is larger than $|c|$, then $k_{2}$ can be large compared with
 $k_{1}$, which means that $\delta \phi_{2}$ can be much more suppressed by expansion
 than $\delta \phi_{1}$ \cite{dyn}. (During matter domination $\delta \phi 
\propto a^{-Re(\eta)}$, where $\eta = (3/2-\sqrt{9/4-4 k})/2$, whilst during inflation 
$\delta \phi 
\propto a^{-Re(\sigma)}$, where $\sigma = (3-\sqrt{9-4 k})/2$ \cite{dyn}.)
 Thus depending on $|c|$
 and $|a_{\lambda}|$, $\delta \phi_{1} 
\gg \delta \phi_{2}$ or $\delta \phi_{1} 
\ll \delta \phi_{2}$ at $H \approx m$ are possible.
 
       As $H$ becomes smaller than $m$, the A-term becomes dominated by 
$A_{\lambda \; o}$. The real direction is determined by the A-term, so if we define 
$\phi_{1}$ and $\phi_{2}$ relative to the real direction then these fields are rotated 
at $H < m$ relative to their directions at $H > m$. (At $H > m$, $\phi_{2}$ is damped to
 zero due to the $A$-term in the potential \cite{dyn}.)
In addition, due to the 
A-term, the coherent oscillations of $\phi_{1}$ and $\phi_{2}$ will have a relative phase
 shift typically of the order of 1, resulting in a baryon asymmetry approximately
 proportional to 
the phase $\theta = \phi_{2\;o}/\phi_{1\;o}$ of the condensate field,
 where $\phi_{1\;o}$ and $\phi_{2\;o}$ are the amplitudes of the coherent oscillations. 
If the $\theta$ between is small compared with one and $\delta \phi_{1} 
\gg \delta \phi_{2}$ at $H \gae m$, then $\delta \phi_{2}$ at $H \approx m$
after this rotation will be given by 
$\delta \phi_{2} \approx \theta \delta \phi_{1}$. 
 This is the smallest initial value of the $\delta \phi_{2}$ relative
 to $\delta \phi_{1}$ possible at $H \approx  m$. This limiting
 case is relevant to the possibility of forming Q-axitons
 rather than $\pm$Q-balls and we will 
focus on it in the following. 

            We can then consider the growth of perturbation until they become non-linear. 
This has been studied analytically for the case of a maximally charged (MAX) $Q=1$ AD
condensate \cite{ks,bbb2,dyn}. However, it has not been
 considered in detail for the case of a non-MAX condensate, for which the amplitude of 
$\phi_{1}$ is
large compared with that of $\phi_{2}$. 
The equations of motion are  
\be{e3} \ddot{\phi}_{i} + 3H \dot{\phi_{i}}- \nabla^{2} \phi_{i} = -m^{2}(1+K)\phi_{i} 
-K m^2 \phi_{i} 
log\left(\frac{\phi_{1}^{2} + \phi_{2}^{2}}{\phi_{o}^{2}} \right) \;\; , \;\;\; i =1,2 \;\;\;  ~.\ee
For $\phi_{1}(t) \gg \phi_{2}(t)$ during most of the oscillation cycle, the equation for
 $\phi_{1}$ will be similar to the case of the 
MAX condensate, except that $\phi_{1}^{2} + \phi_{2}^{2}$ will be varying periodically
 in time 
rather than constant. Thus the equation for the growth of space-dependent perturbations
 of
$\phi_{1}$ will be similar to the MAX condensate and we expect that the $\phi_{1}$ 
perturbations 
will go non-linear ($\delta \phi_{1}/\phi_{1} \gae 1$) essentially as in the case of the 
MAX condensate \cite{ks,bbb2}. The equation
 for perturbations in $\phi_{2}$ will, however, be different, 
since the {\it log} term is dominated by $\phi_{1}^{2}$. 
Even if $\delta \phi_{2} > \phi_{2}(t)$, the equation for perturbations of $\phi_{1}
 (\equiv \phi_{1}(t) + \delta \phi_{1})$ is not significantly altered so long as $\delta
 \phi_{2}/\phi_{1} < 1$. Thus the perturbation in the energy density (which is essentially
 determined by $\phi_{1}$) will not go non-linear until $\delta \phi_{1}/\phi_{1} \gae 1$,
 even if $\delta \phi_{2}/\phi_{2} \gg 1$. 

       We next show that $\delta \phi_{2}/\phi_{1} \lae \delta \phi_{1}/\phi_{1}$ at all times if
 $\delta \phi_{2} \lae \delta \phi_{1}$ initially at $H \approx m$. To prove this we must
 show that
the rate of growth of $\delta \phi_{2}$ is no greater than that of $\delta \phi_{1}$. We
 consider $\phi_{1} \gg \phi_{2}$, as is true for most of the 
oscillation cycle, and expand the equations of motion, 
\be{e3a} \delta \ddot{\phi}_{1} + 3H \delta \dot{\phi_{1}}- \nabla^{2}
 \delta \phi_{1} = -m^{2}(1+3 K)\delta \phi_{1} 
-2 K m^2 \phi_{2} \delta \phi_{2}/\phi_{1} 
 ~, \ee
\be{e3b} \delta \ddot{\phi}_{2} + 3H \delta \dot{\phi_{2}}- \nabla^{2}
 \delta \phi_{2} = -m^{2}(1+K)\delta \phi_{2} 
-2 K m^2 \phi_{2} \delta \phi_{1} /\phi_{1}
 ~. \ee
Since $\theta \delta \phi_{1} = (\phi_{2}/\phi_{1}) \delta \phi_{1} 
\lae \delta \phi_{2}$, 
the magnitude of the $|K|$ dependent term of the $\delta \phi_{2}$ 
equation, which is responsible for the growth of the $\delta \phi_{2}$
 perturbations, is less than that 
of the $\delta \phi_{1}$ equation. 
Thus if $\delta \phi_{2} \lae \delta \phi_{1}$ initially at $H \approx m$ then
 at all times $\delta \phi_{2} \lae \delta \phi_{1}$. In particular, 
 $\delta \phi_{2} \approx \theta \delta \phi_{1}$ is a consistent soultion of the equations. 
 Therefore for $\delta \phi_{2} \lae \delta \phi_{1}$ at the onset of coherent oscillations,
 non-linearity will 
occur only once $\delta \phi_{1}$ goes
 non-linear, at which point the condensate will 
begin to fragment to condensate lumps. 

               In general this means that 
$\delta \phi_{2}/\phi_{2} \gg \theta$ is possible when
 the energy density in condensate goes non-linear ($\delta \phi_{1} \approx \phi_{1}$)
 and the condensate fragments
 i.e. the fluctuation
 in the phase of the condensate
can be large compared with the average phase. In this case
 we expect $\pm$Q-balls to
 form, since the charge density of the non-MAX condensate is proportional to $\phi_{1}
 \phi_{2} \approx \phi_{1}\delta \phi_{2}$ and $\delta \phi_{2}$ is periodic in space with
 positive and negative values.
 However, in the limiting case where $\delta \phi_{2}
 \approx \theta \delta \phi_{1}$
at $H \approx m$ (and $\theta$ is small compared with 1), $\delta \phi_{2}
 \approx \theta \delta \phi_{1}$
is a consistent solution throughout. Therefore
 when the condensate fragments
 we will have $\delta \phi_{2} / \phi_{2} \approx 1$ 
i.e. the fluctuation of the phase of the
 condensate is approximately the same as the
 average phase. In this limiting case we expect that 
Q-axiton formation will be favoured over $\pm$Q-ball formation.
 A full numerical evolution of the perturbations will be
 necessary to determine exactly what happens in this case, but we expect that 
there will be dominant 
formation of Q-axitons together with a few $\pm$Q-balls. 

                    So the question of whether Q-axitons or $\pm$Q-balls form will be 
detemined by the fluctuation in the phase at the onset of condensate fragmentation. This 
will all depend upon the evolution of the perturbations after they leave the horizon during 
inflation, which in turn depends on the parameters of the AD scalar potential. We believe 
that both cases are possible, with $\pm$Q-ball formation being typical but with Q-axiton
 formation as a limiting case. 

\subsection{Numerical simulations and absorption/re-emission of scalar waves} 

           We next consider the classical evolution of the non-linear condensate lumps. For the
case of Q-balls (MAX condensate or non-MAX with $\pm$Q-balls), this is 
straightforward. Non-relativistic Q-balls carrying essentially all the charge form directly
 from the condensate and move apart from each other due to the expansion of the
 Universe until 
their decay rate to quarks is fast enough relative to the expansion rate to allow them to
 evaporate 
away into baryons and neutralinos \cite{bbb2}. For the case of a non-MAX condensate with
 Q-axitons, the non-linear evolution of a single spherically symmetric condensate lump
 appears to show that 
$f_{B} < 1$ once the lump reaches the classically stable Q-axiton state \cite{dyn}. It
 takes some time to 
reach the Q-axiton state, via emission of scalar field waves
 from a pulsating condensate lump \cite{dyn}. However, in realistic models we cannot
 consider a 
single, isolated condensate lump, but instead must consider an ensemble of lumps distributed throughout space.
 This leads to the possibility that scalar waves emitted by 
one relaxing condensate lump might be absorbed by a neighbouring condensate lump. 
(Since the phase of the Q-axitons (ranging roughly from 0 to $2 \theta$)
 on average will be the same
 as the phase $\theta$ of the initial condensate lump, absorption and
 emission of scalar field waves should be relatively efficient and 
 without reflection effects due to large (order $\pi$) phase differences \cite{vilja}.)
As a result it is possible  
that the condensate lumps are maintained in an excited state
 until expansion pulls them far 
enough apart for absorption/re-emission to be no longer
 effective. Only then can the condensate lumps reach their
 classically stable Q-axiton state and 
$f_{B}$ be extracted from lattice simulations \cite{kk} of condensate collapse. 

      The condition for absorption/re-emission to be important can be estimated as follows. 
We can consider a "lattice" of condensate lumps to form just after condensate collapse.
These lumps will pulsate about their stable Q-axiton configuration, emitting scalar field
 waves as they do \cite{dyn}. For 
simplicity, we will replace the discrete condensate lumps with a continuous number density,
 in order to estimate the number of condensate lumps encountered by an expanding scalar
 wavefront. Suppose the condensate collapses to form condensate
 lumps at $t_{o}$, $H_{o}$. Consider a 
scalar field pulse emitted by a lump at a later time
 $t_{i}$. The condensate lump density is
\be{e4} n_{Q}(t) = n_{Q\;o} \left(\frac{a_{o}}{a}\right)^{3}    ~,\ee
where $n_{Q \; o} = n_{Q}(t_{o})$. 
The charge absorbed by the surrounding condensate lumps may
 be estimated by considering the area of the 
outgoing wave removed by encountering other lumps. The lumps have area 
\be{e5}  A_{l} \approx 4 \pi r_{l}^{2}       ~,\ee
where $r_{l}$ is the radius of the lump. 
The charge density of the outgoing wavefront of radius $r$ is 
\be{e6} \sigma_{Q} = \frac{Q_{W}}{4 \pi r^{2}}    ~,\ee
where $Q_{W}$ is the total charge carried by the wavefront.
Thus the charge lost upon encountering a condensate lump
 is $\sigma_{Q} A_{l}$. The
charge absorbed from a wave emitted at $t_{i}$ is therefore
\be{e7} \Delta Q = \int_{t_{i}}^{t} n_{Q}(t) \sigma_{Q} A_{l} 
4 \pi r^{2}(t) v(t) dt = n_{Q\;i} Q_{W} A_{l} v_{o} H^{-1}_{i} 
\int_{a_{i}}^{a} \left(\frac{a_{i}}{a}\right)^{7/2} a_{i}^{-1} da    
~,\ee
where the velocity $v(t)$ of the scalar field
 waves is initially set by a 
wavelength of the the order of the Q-axiton
 radius, corresponding to momentum 
$k \approx 2 \pi |K|^{1/2} m$ \cite{dyn}, which implies that 
the scalar field waves are initially mildly 
relativistic, $ v(t_{i}) = v_{o} \approx 1$, and that subsequently
$v = v_{o}(a_{i}/a)$. Therefore the charge absorbed is
\be{e8} \frac{\Delta Q}{Q_{W}}  = \frac{2}{5} \frac{n_{Q\;i} A_{l} v_{o}}{H_{i}}
\left(1 - \left(\frac{a_{i}}{a}\right)^{5/2}\right)      ~.\ee 
Thus for $a_{i}/a > 1$ the charge absorbed is given by
\be{e9} \frac{\Delta Q}{Q_{W}} \approx \frac{2 A_{l}
 n_{Q\;i} v_{o}}{5 H_{i}}    ~.\ee
The condition for total absorption of the outgoing wave is
 $\frac{\Delta Q}{Q_{W}} 
> 1$.  In terms of $H_{o}$ and $a_{o}$, 
$H_{i} = H_{o} \left(\frac{a_{o}}{a_{i}}\right)^{3/2}$. 
When the condensate lumps form, the radius of the spherically
 symmetric condensate lumps is approximately the same as the 
spacing between the lumps, $r_{l} \approx (|K|^{1/2} m)^{-1}$ \cite{bbb2,dyn}. 
Condensate lumps form at $H_{o} \approx
(2 |K| m)/\alpha(\lambda_{o})$ where $\alpha(\lambda_{o})
 \approx 30$ \cite{bbb2}. Thus from \eq{e9} the 
condition for absorption by condensate lumps to be negligible
 is that the expansion factor must satisfy $a > a_{c}$, where
\be{e10} \frac{a_{c}}{a_{o}} = \left(\frac{3 v_{o}}{5}\right)^{2/3} 
\frac{\alpha^{2/3}}{|K|^{1/3}}       ~.\ee
This must be satisfied in order to be certain that a numerical simulation of
 condensate fragmentation has allowed the 
condensate lumps to reach their Q-axiton equilibrium
 configuration, in which case $f_{B}$ can be 
extracted. In a recent simulation \cite{kk}
of condensate fragmentation on the lattice, Kasuya and Kawasaki found that generally
 $f_{B} = 1$ to 
a good approximation. 
Their simulation runs from initial time
 $t_{o} = 5 \times 10^{3} m^{-1}$ to $t_{f} = 
10^{5}m^{-1}$ and for values of
 $|K| = 0.01,\;0.05,\;0.1$. The scale factor at the end of
 their simulation, $a_{f}$, is thus 
\be{e11}    \frac{a_{f}}{a_{o}} = 
\frac{\left(1 + \frac{3}{2}H_{o}t_{f}\right)^{2/3}}{
\left(1 + \frac{3}{2}H_{o}t_{o}\right)^{2/3}} \approx 
\left(\frac{t_{f}}{t_{i}}\right)^{2/3} = 6.9    ~.\ee
On the other hand, the values of $a_{c}/a_{o}$ for
 $|K| = (0.01,\;0.05,\;0.1)$ are (31.7, 18.5, 14.3). Thus 
the simulation is well within the time scale during which 
absorption/re-emission will be efficient. This might account for the large value of 
$f_{B}$ obtained in \cite{kk} as compared with isolated spherically
 symmetric lump evolution. In order to ascertain
 whether $f_{B} = 1$ is 
correct, or whether a $|K|$ and $Q$ dependent value of $f_{B}$
 can occur as suggested
by single lump evolution \cite{dyn}, a longer simulation may be necessary.

   The simulation in \cite{kk} also shows that for smaller angular velocities of the complex 
condensate field (i.e. amplitude of $\phi_{2}$ small compared
 with $\phi_{1}$), $\pm$Q-ball formation becomes important. This is
 consistent with fluctuations in the phase becoming larger than the average
 phase ($\phi_{2}/\phi_{1}$) when the condensate fragments, assuming that the initial
perturbation of the condensate is unaltered as the angular velocity ($\phi_{2}$) is reduced. 

\section{Evolution of the Q-axiton} 

        We next consider if and how a Q-axiton relaxes to the lower energy stable
Q-ball. In numerical simulations \cite{dyn} the Q-axiton is stable on time scales 
much larger than $m^{-1}$; the attractive force between the scalars in the 
condensate is balanced by the gradient energy\begin{footnote}{Somewhat similar long-lived "pseudo-breather states" are also observed in $3+1$ Sine-Gordon theory, which may represent long-lived pionic states that could be produced in heavy-ion collisions \cite{hh}.}\end{footnote}
On the other hand, it has been shown that 
there is no stable spherically symmetric breather soliton (such as the Q-axiton) in $3+1$ 
dimensions \cite{kt}. Thus we expect that the Q-axiton will evolve to a lower energy 
state, most likely a Q-ball. It is not apparent from the existing numerical simulation how it
 loses energy, but from the equations of motion we can get some idea of the process. In the 
equations of motion for $\phi_{1}$ and $\phi_{2}$ the argument of the $log$ term, \eq{e3},
 $\phi_{1}^{2} + \phi_{2}^{2}$, is varying periodically in time. This is contrast with the
 case of the Q-ball, for which the amplitudes of $\phi_{1}$ and $\phi_{2}$ are the same
 such that $\phi_{1}^{2} + \phi_{2}^{2}$ is time-independent. It is this property that
 allows one to factor out the time dependence in the Q-ball solution, $\Phi(r,t) \sim
 \phi(r)e^{i\omega t}$. For the Q-axiton we cannot factor out the time dependence, 
 so we expect the magnitude of the Q-axiton field to be varying periodically in time
 on a time scale $\approx m^{-1}$, albeit very slightly based on the numerical results \cite{dyn}.
 Thus we expect to find that the Q-axiton has a "quivering" solution. This will then result
 in the emission of scalar field waves carrying energy and perhaps charge. The Q-axiton
 could then either evolve to a Q-ball of the same charge or could even, in the opposite limit,
 completely evaporate away. However, since the rate at which energy is being lost by the
 Q-axiton is very small on the dynamical time scale of the scalar field, $m^{-1}$, and since
 the lowest possible energy final state would be to have all the charge in the
 form of a Q-ball, it is likely that
 this will be the final state of the process.
If this picture is correct then
 although the Q-axiton appears to be stable on relatively long time scales compared with 
$m^{-1}$ \cite{dyn}, on time scales corresponding to the inverse Hubble parameter at the
 time of Q-ball decay it is likely that the Q-axiton will have evolved to the corresponding
 Q-ball. It remains to be shown that Q-axitons do indeed evolve to Q-balls and that the
 time scale for Q-axiton evolution is indeed smaller than that of Q-ball decay. 

              The above process is purely classical in nature. It is also possible that the
 excess energy in the Q-axiton could be released via annihilations of the scalars. We can
 consider the Q-axiton to be, to a first approximation, a superposition of two coherently
 oscillating scalar fields $\phi_{1}$ and $\phi_{2}$. The annihilation rate is 
$n_{i}^{-1}dn_{i}/dt \propto n_{i}\;\;(i=1,2)$. Therefore since $n_{1} \approx m
 \phi_{1}^{2} \gg n_{2} \approx  m \phi_{2}^{2} $, the rate at which the $\phi_{1}$
 amplitude decreases will be much larger than that of the $\phi_{2}$ amplitude. Therefore 
$\phi_{1\;o} \rightarrow \phi_{2\;o}$ i.e. the Q-axiton tends towards a Q-ball. The 
final Q-ball charge to the initial Q-axiton charge will be $Q_{f}/Q_{i} =
 \phi_{2\;f}/\phi_{1\;i} \leq \phi_{2\;i}/\phi_{1\;i} = Q$, the charge of the initial AD condensate. Thus if annihilations
 are 
able to allow the Q-axiton to evolve to a Q-ball before the Q-axiton decays and before the
 neutralinos freeze out of chemical
 equilibrium, then $f_{B} \rightarrow  Q\;f_{B\; cl}$, where $f_{B\;cl}$ is the
 classical value of $f_{B}$ for the Q-axiton. 

            We next consider the decay and annihilation of the scalars in the Q-axiton. Q-balls
 and Q-axitons from AD condensate fragmentation have a
 thick wall profile, 
well-described by a Gaussian profile $\phi(r) = \phi(0) e^{-r^{2}/R^{2}}$, where 
$R \approx   (|K|^{1/2}m)^{-1}$ is the Q-ball/Q-axiton radius \cite{bbb2,kk}. 
Q-ball decay was originally analysed for the case of a thin-wall Q-ball with scalars
 decaying to fermion pairs \cite{qd}. The decay rate was found to be 
\be{d1}   \left(\frac{dB}{dt}\right)_{fermion} \lae  \frac{\omega^{3} A}{192 \pi^{2}}   ~\ee
where $\omega \approx m$ and $A$ is the area of the Q-ball. This is based on scalar decay
 filling the phase space for fermions of energy $m/2$ throughout the Q-ball, 
resulting in Pauli blocking of
 further decay and a Q-ball decay rate which is proportional to
 $A$ since fermions can escape at the surface. In the case of the thick walled Q-ball in the MSSM there
is no equivalent estimate of the decay rate to fermion pairs. However, we expect the order
 of magnitude to be similar to the thin-wall case for a given $\phi(0)$, since the decay rate
 is simply determined by completely filling the fermion phase space within the Q-ball
 volume. In addition, in the case of Q-balls in the MSSM it is possible 
to have decays to pairs of scalars and gauge bosons, which are not suppressed by Pauli
 blocking. In \cite{bbb2} we considered the decay rate to scalars for the case of a thick walled 
Q-ball. It was shown that most of the Q-ball decay came from scalar decays occuring 
in a region of width $\delta r \approx R^{2}/4 r_{*}$ around 
 $r_{*} = \gamma R$ (where $\gamma = ln^{1/2}(g \phi(0)/m) \approx 5$), where tree
 level decay
 to pairs of particles coupling to $\phi$, of mass $g \phi(r)$, becomes kinematically possible. 
The resulting Q-ball decay rate was found to be enhanced by a factor up to $10^{3}$ 
compared with the thin-wall fermion decay rate, but is still proportional to $A$. 

         We assume that the AD scalar decay rate to pairs of bosons or fermions has the form 
$${ \Gamma(r) \approx \frac{\alpha^{2} m^{3}}{\phi^{2}}, \;\;\;\;  g\phi > m  }$$
\be{d2}    \approx \alpha m , \;\;\;\;  g\phi < m  ~,\ee
(which assumes the largest possible decay rate via exchange of heavy particles 
once $g \phi > m$). We first show that in general Pauli blocking/Bose enhancement effects
 are likely to be significant in Q-ball decay. The maximum number density of particles
 before the
 occupation number of states becomes greater than 1 is given by
\be{d3}  n_{max} \approx \frac{p^{2} \Delta p}{\pi^{2}}     ~,\ee
where $p \equiv | \vec{p}| = m/2$ is the momentum of the final state particles and $\Delta
 p$ is the spread in momentum of the final state particles. Since the AD scalars in the
 Q-axiton are confined to be within a radius $R^{-1}$ we expect $\Delta p
 \approx R^{-1}$ and so 
\be{d4} n_{max} \approx \frac{|K|^{1/2} m^{3}}{4 \pi^{2}}     ~.\ee
Pauli blocking/Bose enhancement effects will be important if the density of particles from
 AD scalar decays is greater than $n_{max}$. Relativistic particles can escape from the
 region of the thick-walled Q-ball $\delta r$ around $r_{*}$, where most of the decays 
occur, in a time $\tau \approx \delta r \approx R/20$. Thus the number density of particles
 from AD scalar decay is
\be{d5} n_{decay} \approx n \Gamma_{decay} \tau \approx 
 \frac{4 \pi m^{3}}{20 |K|^{1/2}}     ~, \ee
where we have used $n \approx m \phi^{2}(r)$ and $\phi^{2}(r_{*}) \approx 
m^{2}/g^{2}$. 
Therefore
\be{d6}   \frac{n_{decay}}{n_{max}} \approx \frac{4 \pi^{3}}{5 |K|}  ~.\ee
This is typically large compared with 1, since $|K| \approx 0.01 - 0.1$ in the MSSM
 \cite{asko}. Therefore Pauli blocking or Bose enhancement effects are likely to be
 significant in the decay of AD scalars in the Q-axiton. The result of this is that all the
 scalars in the region $\delta r$ around $r_{*}$ will rapidly decay. These will be
 replenished by the Q-axiton reconfiguring itself on the dynamical time scale $m^{-1}$. In
 the absence of Bose enhanced decay, in a time $m^{-1}$ the fraction of the scalars in the
 region $\delta r$ which decay is $\Gamma_{decay} m^{-1} \approx \alpha$. Therefore
 the enhancement in the decay rate is of the order of $\alpha^{-1}$, which will increase
 the decay temperature of the Q-axiton by $\alpha^{-1/2}$.

     We next compare the decay and annihilation rates. Most of the decays occur around 
$r \approx r_{*}$ via tree level decays of the AD scalars \cite{bbb2}.
 We first show that AD scalars in the Q-axiton
can annihilate at tree-level to pairs of gauge bosons but cannot decay to pairs of gauge
 bosons. To do this we consider a simple $U(1)$ gauge interaction with 
a pair of "quark" superfields $Q_{1}$ and $Q_{-1}$, with $U(1)$ charges 
$\pm 1$.  
The terms of the Lagrangian associated with the gauge interaction are then 
$${ {\cal L}  = D_{\mu}Q_{1}^{\dagger}D^{\mu}Q_{1} 
+ D_{\mu}Q_{-1}^{\dagger}D^{\mu}Q_{-1}
+ (ig  Q_{1}^{\dagger}\overline{\lambda} \psi_{Q_{1}} 
- ig Q_{-1}^{\dagger}\overline{\lambda}  \psi_{Q_{-1}} + h.c) }$$
\be{l1} + \frac{g^{2}}{2} \left( |Q_{1}|^{2} - |Q_{-1}|^{2}\right)^{2}      ~.\ee
We can rewrite this in terms of a flat direction scalar $\Phi$ and its orthogonal 
partner $\Theta$, 
\be{d7} \Phi = \frac{1}{\sqrt{2}} (Q_{1} +  Q_{-1})     ~,\ee
\be{d8} \Theta = \frac{1}{\sqrt{2}} (-Q_{1} +  Q_{-1})     ~.\ee
The kinetic terms in ${\cal L}$ are then 
\be{d9}  (\partial_{\mu}\Phi)^{\dagger}\partial^{\mu}\Phi
+ (\partial_{\mu}\Theta)^{\dagger}\partial^{\mu}\Theta 
+ g^{2} A^{\mu}A_{\mu}( \Phi^{\dagger}\Phi + 
\Theta^{\dagger}\Theta) 
-g A^{\mu}(i \Phi \partial_{\mu}\Theta^{\dagger} + 
i \Theta \partial_{\mu}\Phi^{\dagger} + h.c.)    ~.\ee
From this we see that on introducing an expectation value for $Re(\Phi)$,
representing the Q-axiton field,
the Goldstone boson corresponds to $Im(\Theta)$. 
Thus pairs of Q-axiton scalars couple to the massive gauge bosons but there is no 
linear coupling of $Re(\Phi)$ to gauge bosons once the Goldstone boson is rotated away.
 As a result, we can have Q-axiton scalar annihilations to pairs of gauge boson
but no decays to pairs of gauge bosons. The scalars can, however, decay to 
gaugino-quark pairs via the interaction
\be{d10}   i g  Q_{1} \overline{\psi}_{Q_{1}} \lambda
- i g  Q_{-1} \overline{\psi}_{Q_{-1}} \lambda + h.c. \rightarrow
-i g \Phi \overline{\psi}_{\Theta} \lambda + h.c.  ~,\ee
with decay rate
\be{d11}   \Gamma_{decay} \approx \alpha m    ~,\ee
where $\alpha = g^{2}/4 \pi$.
Pairs of AD scalars can annihilate to gauge boson pairs, with annihilation cross-section 
\be{d12} \sigma_{ann} \approx \frac{\alpha^{2}}{m^{2}}   ~.\ee
Thus the annihilation rate is 
\be{d13} \Gamma_{ann} = n \sigma_{ann} \approx m \phi(r^{*})^{2} \sigma_{ann} 
\approx \frac{\alpha m }{4 \pi}   ~,\ee
where $n$ is the number density of AD scalars.
From this we expect that the annihilation rate to gauge bosons will be no larger than
  the decay rate to gaugino-squark pairs.
The only way to have a significant suppression of the decay rate relative to the annihilation
 rate would be if the decay to gauginos was kinematically suppressed such that 
the decay occurs only to quarks and squarks via small Yukawa couplings. This would
 depend on the mass of the
 AD scalar relative to the gauginos and the generational structure of the mixture of 
squarks forming the AD scalar. Bose enhancement of the annihilation rate to gauge bosons
 and Pauli blocking of decay rate to gauginos could also help to enhance annihilations
 relative to decays.

        Therefore it is possible that in the limiting case where the non-MAX condensate
 fragments to Q-axitons rather than $\pm$Q-balls, the Q-axiton can evolve to a Q-ball
 before the Q-balls decay,
 either by classical evolution from the quasi-stable Q-axiton state or, less likely, by scalar
 annihilations
 dominating over decays. The value of $f_{B}$ when the Q-balls decay is particularly
 small in the annihilation case, which could be important in allowing a wide range of
 neutralino masses
 to be consistent with the baryon to dark matter ratio. However, a more precise analytical
  and numerical study will be required to show exactly how the Q-axiton evolves to a
 Q-ball. 

\section{Consequences for the Dark Matter and the Baryon to Dark Matter Ratio} 

\subsection{Neutralinos directly from Q-ball decay}

            As discussed in the Introduction, a major motivation for B-ball Baryogenesis is the
similarity, to within an order of magnitude, of the {\it number} densities of baryon and
 dark
 matter particles for the case where dark matter particles have masses of the order of
 $m_{W}$. Q-balls from a $d=6$ AD condensate are expected to decay at a temperature
 $T_{d} \approx 1 \MeV - 1 \GeV$ \cite{bbb2}.
Since the scalars in the Q-ball (or Q-axiton) are $R_{p}$-odd (essentially squarks and
 sleptons) they will produce
one LSP neutralino per scalar. For the case of Q-balls from a MAX condensate,
assuming that there is no subsequent annihilation of the neutralinos from Q-ball decay,
the number density of neutralinos will  be related to the baryon number density by
\be{e14} n_{\chi} = 3 f_{B} n_{B}         ~,\ee
or equivalently
\be{e15} \Omega_{\chi} = 3 f_{B} \left(\frac{m_{\chi}}{m_{n}}\right)  \Omega_{B}    ~.\ee
Nucleosynthesis constrains $\Omega_{B}$; we consider two possible ranges,
\be{e16} 0.0048 \lae \Omega_{B}h^{2} \lae 0.013    ~\ee
from "reasonable" bounds on primordial element abundances \cite{bbb2,sarkar} and
\be{e17} 0.004 \lae \Omega_{B}h^{2} \lae 0.023    ~\ee
 from conservative bounds \cite{kk,olivens}. With $h = 0.6-0.8$ \cite{freedman,jrev} we 
obtain upper bounds on the LSP mass neutralino:
\be{e18} m_{\chi} < 17.6 \; f_{B}^{-1} \left(\frac{\Omega_{\chi}}{0.4}\right) 
\left(\frac{h}{0.8}\right)^{2} \GeV     ~\ee
from the reasonable nucleosynthesis bounds and 
\be{e19} m_{\chi} < 20.8 \; f_{B}^{-1} \left(\frac{\Omega_{\chi}}{0.4}\right) 
\left(\frac{h}{0.8}\right)^{2}  \GeV    ~\ee
from the conservative bounds.

          For the case of the MSSM, present experimental bounds on the LSP 
neutralino mass depend upon the theoretical assumptions made. The direct experimental 
lower limit on the neutralino mass from ALEPH is $m_{\chi} > 32.3 \GeV$ \cite{hutch},
 which requires that $f_{B} \lae 0.64$, assuming the conservative nucleosynthesis limit. 
If one evolves the renormalization group equations assuming universal A-terms and
 gaugino masses at the unification scale then LEP and Tevatron constraints imply that
 $m_{\chi} > 46 \GeV$ ($m_{\chi} > 51 \GeV$ with universal scalar masses also)
 \cite{oliveetc}, which requires that $f_{B} \lae 0.45$ ($f_{B} \lae 0.41$ with universal
 scalar masses), assuming the conservative nucleosynthesis limits. Thus the case of Q-balls
 with $f_{B} \approx 1$ (expected for the MAX condensate), is not compatible with
 MSSM neutralinos.  

            If we consider a non-MAX condensate then we can either form $\pm$Q-balls or,
 as a limiting case, Q-axitons. If the Q-axitons cannot evolve to Q-balls before the 
neutralinos freeze-out then the extra energy in the Q-axiton relative to a Q-ball of the
 same charge will result in more neutralinos being produced in the decay, such that 
$f_{B} \rightarrow f_{B}/Q$ in the $m_{\chi}$ upper bounds. The same correction is
 necessary in $\pm$Q-ball decay. (So as far as the baryon to dark matter ratio is concerned
 there is no real difference between Q-axitons and $\pm$Q-balls in this case.) So for
 $Q<1$ the constraints on $f_{B}$ are
 correspondingly tighter, further disfavouring Q-ball decay as an explaination 
of the baryon to dark matter ratio in the MSSM. (For the $\pm$Q-ball case, $f_{B} = 1$
is expected, whereas for the Q-axiton $f_{B}$ is estimated to be as small as 0.3 for
 $Q \approx 0.01$ \cite{dyn}. However, even in this case
$f_{B}/Q$ is too large to be compatible with MSSM neutralinos.)

    Thus the only way the MSSM can be compatible with direct Q-ball decay to 
neutralinos as an explaination of the baryon to dark matter ratio is if we have the limiting
 case where Q-axitons form from a non-MAX condensate and evolve to Q-balls before the
 neutralinos freeze-out. 
In this case, for small condensate charge $Q$, $f_{B}$ could be small enough to allow
 dark matter MSSM neutralinos compatible with experiment to come directly from Q-ball
 decay, especially in the 
case where annihilations are responsible for the evolution of the Q-axitons to 
Q-balls.  

        If we take the view that the baryon to dark matter ratio is indeed due to decaying 
Q-balls then, modulo the possibility of Q-axiton formation and evolution, we must
 consider an alternative
 to MSSM LSP neutralinos. A natural possibility is to consider the NMSSM with 
a mostly singlino LSP \cite{sing,stephan,klap}. With $f_{B} \approx 1$
 and $Q \leq 1$, Q-ball decay imples that the
LSP must have a  mass less than about $20 \GeV$. Current experimental bounds
 combined with renormalization group 
evolution indicate a lower bound of 3-5 $\GeV$ on the singlino LSP mass \cite{klap}. 
(Singlinos may be excluded by vacuum stability considerations, but this depends on the
 universality conditions at the unification scale \cite{stephan}). 

                So, given that 
$f_{B}/Q \gae 1$ is likely when the Q-balls decay, Q-ball decay combined with 
the baryon to dark matter number ratio and experimental bounds on neutralino masses
 rules out the MSSM but is compatible with the NMSSM with a mostly singlino LSP of
 mass less than about 20$\GeV$. This may be regarded as a prediction of the B-ball
 Baryogenesis explaination of the baryon to dark matter ratio. 
 
\subsection{Effect of annihilations}

        In the above it has been assumed that the neutralinos from Q-ball decay do not
 annihilate. However, if a large enough density of neutralinos is produced in 
Q-ball decay, they will annihilate with each other. In this case the number density 
of LSP neutralinos is determined by the annihilation cross-section and is not directly 
related to the baryon number density. It will also be spatially constant, ruling out the
 possibility of 
neutralino isocurvature fluctuations correlated with the baryon number \cite{iso}. 

      The upper limit on the number of LSPs at a given temperature is \cite{bbb2}
\be{e20} n_{limit}(T) = \left(\frac{H}{<\sigma v>_{ann}}\right)_{T}      ~,\ee
where $\sigma$ is the annihilation cross-section, $v$ is the relative velocity of the
 neutralinos and $<...>$ denotes the thermal average.
For neutralinos $<\sigma v>_{ann} = a + bT/m_{\chi}$, where $a$ and $b$ are determined 
by the parameters of the SUSY model \cite{ellisdm}. For the case of light neutralinos
$m_{\chi} < m_{W}$, the $a$ term is negligible and the annihilation cross-section is 
$b$ dominated, in which case $n_{limit} \propto g(T)^{1/2} T$. Therefore 
\be{e21} n_{limit}(T) = \left(\frac{g(T_{f})}{g(T)}\right)^{1/2} 
\left(\frac{T_{f}}{T}\right)^{3} n_{relic}(T)      ~,\ee
where
\be{e22} n_{relic}(T) = \left(\frac{g(T)}{g(T_{f})}\right) 
\left(\frac{T}{T_{f}}\right)^{3} \left(\frac{H}{<\sigma v>_{ann}}\right)_{T_{f}}       ~\ee
is the thermal relic neutralino density and $T_{f}$ is the neutralino freeze-out temperature
 (typically $T_{f} \approx m_{\chi}/20$).
Thus in order to account for the baryon to dark matter ratio the LSP density 
from Q-ball decay must satisfy
\be{e23}   n_{relic}(T_{d}) < n_{LSP}(T_{d}) < n_{limit}(T_{d})    ~.\ee
In general 
\be{e24}  n_{LSP}(T_{d})  = \frac{3 f_{B} n_{B}}{Q}     ~.\ee
Thus the condition 
$n_{LSP}(T_{d}) < n_{limit}(T_{d}) $ implies that \cite{bbb2}
\be{e25}  T_{d} < T_{d\;c} \approx 1.2 \left(\frac{Q}{f_{B}}\right)^{1/2}
\left(\frac{100}{g(T_{f})}\right)^{1/2}
\left(\frac{10^{-10}}{\eta_{B}}\right)^{1/2}
\left(\frac{m_{\chi}}{100 \GeV}\right)
~.\ee
Since the decay temperature is estimated to be between $1 \MeV$ and $1 \GeV$
 \cite{bbb2}, for reasonable values of $Q$ and 
$f_{B}$ (say $Q \gae 0.1$) we expect that it is quite natural for this to be satisfied,
 especially for the case of a maximally charged condensate with $Q = f_{B} = 1$.
 However, if $T_{d} 
> T_{d\;c}$ then the LSP density is given by $n_{limit}(T_{d})$, independently of the 
details of Q-ball decay,
\be{e26} n_{limit}(T_{d}) = \left(\frac{T_{d\;c}}{T_{d}}\right)^{2} n_{LSP}(T_{d})  
~.\ee  
Although the direct connection between the baryon number and dark matter number
 density is lost, the neutralino dark matter will still be non-thermal in nature. 

\section{Non-Spherical Collapse and Velocity Effects} 

       In the above discussion we have been considering values of $f_{B}$ based on 
the spherically symmetric collapse of a single condensate lump and the evolution of an
 ensemble of static lumps being pulled apart from each other by the expansion of the 
Universe.  However, it has been suggested that the effects of the velocity 
of condensate lumps after fragmentation \cite{enqvist} and the non-spherical collapse of
 condensate lumps \cite{kk,enqvist} may have very significant effects on the evolution of
 the ensemble of condensate lumps. In a recent numerical simulation \cite{enqvist} it was
 observed that after condensate fragmentation the lumps initially have a mildly relativistic
 velocity and can undergo collisions. 
(However, the lump velocity may depend upon the assumed form of the
 initial perturbation at $H \approx m$. For example, a standing wave perturbation may
 result in static lumps.)
In addition, the large lumps which initially formed subsequently fragmented
 into $\pm$Q-balls. One question here is whether the formation of 
$\pm$Q-balls is due to the dynamics of condensate collapse or due to the linear
 growth
 of the perturbations of the charge density, as discussed in Section 2? 
As well as mildly relativistic large Q-balls, smaller relativistic Q-balls
 were formed
 during condensate collapse. These, it is suggested, can effectively thermalize
 the distribution, allowing an analytical treatment \cite{enqvist}.  The possibility of 
forming a significant number of small Q-balls in condensate fragmentation could have a 
significant effect on $f_{B}$. These could be emitted during the evolution of a
 non-spherical condensate lump, either due to being initially non-spherical after formation
 or due to the fusion of two colliding condensate lumps; for example, in \cite{kk} it was
 shown that in a $1+1$ dimensional simulation of Q-ball fusion, 7$\%$ of the charge was 
emitted in the form of very small Q-balls (or possibly scalar field waves, as the lattice
 simulation cannot distinguish these cases \cite{kk}). Since the rate of decay of Q-balls is
 proportional to $Q^{-1}$ and the decay temperature $T_{d} \propto Q^{-1/2}$, 
it is possible that very small Q-balls could decay before the neutralinos  freeze out of
 chemical equilibrium, $T_{d} > T_{f}$. In this case, since the neutralinos from Q-ball
 decay thermalize to become part of the thermal relic density and so are not related to the
 baryon number from the decaying Q-balls, the charge in small Q-balls is effectively
 removed from the Q-ball
 ensemble before the larger Q-balls decay, reducing $f_{B}$. So the tendency of 
non-spherical lumps and velocity on the formation and evolution of condensate lumps is to
 reduce $f_{B}$ as compared with the idealized spherically symmetric static case. 

\section{Conclusions}

         We have considered a number of issues connected with the fragmentation of 
the Affleck-Dine condensate and the formation of late-decaying Q-balls, in particular 
the fraction of the baryon number in the Q-balls when they decay, $f_{B}$. For a
 non-maximally
 charged condensate we expect that either $\pm$Q-balls or Q-axitons
will form, depending on the initial fluctuation of the phase of the condensate at $H \approx
 m$.  We showed that in the limiting case where Q-axitons form numerical simulations of
 condensate
 fragmentation must be run for a sufficiently long time to ensure that
 absorption/re-emission of 
scalar field waves does not maintain the condensate lumps far from their classically stable
 Q-axiton state,  so allowing the value of $f_{B}$ to be extracted. 

       A major motivation for B-ball baryogenesis is a natural explaination for the baryon to 
dark matter number density ratio. 
 In general, Q-ball formation with $f_{B} = 1$ from fragmentation of a MAX condensate and 
 $\pm$Q-ball formation from a non-MAX condensate produces too many neutralinos per
 baryon
 number to be consistent with experimental bounds on MSSM neutralino masses if we
 wish to account for the baryon to dark matter particle number ratio via Q-ball decay. The
 only 
possibility is to have the limiting case of Q-axiton formation from a non-MAX condensate,
 with evolution of 
the Q-axitons to Q-balls before neutralinos freeze out of chemical equilibrium, such that
 $f_{B} \ll 1$ is possible. This is particularly possible if the Q-axiton evolves to a Q-ball
 via annihilations.  

        It is also possible that more complex features of AD condensate fragmentation which
 can only be studied numerically, such as the formation and evolution of non-spherically
 symmetric condensate lumps and collisons of moving lumps, might reduce the value of
 $f_{B}$, but it remains to be seen if these effects can reduce it sufficiently to accomodate
 MSSM neutralinos in the typical case of Q-ball formation from a MAX or
$\pm$Q-ball formation from a non-MAX condensate. 

     In the typical case with $f_{B}/Q \gae 1$ (rather than the limiting case of Q-axiton
 formation followed by evolution to Q-balls), the LSP neutralino mass from direct 
Q-ball decay must be less than about 20GeV. This is too large to be consistent with 
MSSM neutralinos ($m_{\chi} >  32.2\GeV$ experimentally), but is consistent with singlino dark matter in the NMSSM. Thus the B-ball baryogenesis explaination of the 
baryon to dark matter ratio
strongly suggests that dark matter is non-thermal and that the LSP should be associated
 with an extension of the MSSM with an experimentally allowed light LSP, 
the NMSSM being the most obvious possibility. 
In addition, depending on details of the inflation
 model, there may be observable isocurvature density fluctuations due to the transfer of
 baryon isocurvature fluctuations to neutralinos via Q-ball decay \cite{iso}. 
Although all the evidence for B-ball baryogenesis is circumstantial, and each element 
could have an alternative explaination, if it turns out that the NMSSM is 
 realized in nature with a light singlino LSP, if the LSP dark matter density is 
inconsistent with a thermal relic density and if isocurvature density perturbations 
are observed, then $d=6$ B-ball baryogenesis would become a favoured scenario for 
the simultaneous origin of both the baryon number and dark matter density. 

     Finally, it should be emphasized that much work remains to be done to clarify the
 physics of unstable AD condensates and B-ball Baryogenesis. 
Given that $d=6$ AD baryogenesis is a serious candidate
 for the origin of the baryon asymmetry in SUSY models and that late-decaying
 Q-balls could naturally account for the baryon to dark matter ratio, it 
is important to develop in detail the physics of AD condensate fragmentation in the early
 Universe.

\section*{Acknowldgements}

         The author would like to thank Kari Enqvist for a most enjoyable
 collaboration and Tim Jones and Keith Olive for their support over the years.

\end{document}